\documentclass[a4paper,11pt]{article}
\usepackage{amsfonts}
\usepackage[dvips]{epsfig, graphics}

\newcommand{\ti}{\times}
\newcommand{\gsim}
{\;\raisebox{-.3em}{$\stackrel{\displaystyle >}{\sim}$}\;}

 \newlength{\wth}
 \newcommand{\twographs}[2]{%
 \unitlength=1.1in
 \begin{picture}(5.8,2.6)
 \put(0,0){\epsfig{file=#1.eps, width=1.1\wth}}
 \put(2.3,0){\epsfig{file=#2.eps, width=1.1\wth}} 
 \put(0,2.1){(a)}
 \put(2.3,2.1){(b)}
 \end{picture}
}

\newcommand{\threegraphs}[3]{%
 \unitlength=1.1in
 \begin{picture}(6.,5.2)
 \put(1.1,0){\epsfig{file=#3, width=1.2\wth}}
 \put(-0.2,2.6){\epsfig{file=#1, width=1.2\wth}}
 \put(2.2,2.6){\epsfig{file=#2, width=1.2\wth}}
 \put(1.1,2.1){(c)}
 \put(0,4.7){(a)}
 \put(2.2, 4.7){(b)}
\end{picture}}

\begin{document}

\title{\begin{flushright} \vspace{-2cm}
{\small DAMTP-2005-119\\ \vspace{-0.35cm}
hep-th/0512081} \end{flushright}
\vspace{3cm}
{\bf Low-Energy Supersymmetry Breaking from String Flux
  Compactifications: Benchmark Scenarios}
\author{}
\date{}
}
\maketitle

\begin{center}
Benjamin C. Allanach, Fernando Quevedo and Kerim Suruliz

\vspace{0.5cm}
\emph{DAMTP, Centre for Mathematical Sciences,} \\
\emph{Wilberforce Road, Cambridge, CB3 0WA, UK} \\
\end{center}

\begin{abstract}

\noindent Soft supersymmetry breaking terms were recently derived for 
 type IIB string
flux compactifications with all moduli stabilised. Depending on the choice of the discrete
input parameters of the compactification such as fluxes and ranks of hidden gauge groups, the string scale
was found to have any value between the TeV and GUT scales. We study
 the 
phenomenological implications of
these compactifications at low energy. Three realistic 
scenarios can be identified depending on whether the Standard Model
 lies 
on D3 or D7 branes and
on the value of the string scale. For the MSSM on D7 branes
and the string scale between $10^{12}$ GeV and $10^{17}$ GeV we find
that the LSP is a neutralino, while 
for lower scales it is the stop. At the GUT scale
the results 
of the fluxed MSSM are reproduced, but now with all moduli stabilised.
For the MSSM on D3 branes
we identify two realistic scenarios. The first one corresponds to 
an intermediate string scale version of split
supersymmetry. The second is a 
stringy mSUGRA scenario. This requires tuning of the flux parameters
 to obtain the GUT scale. Phenomenological constraints from dark
 matter, $(g-2)_\mu$ and $BR(b\rightarrow s\gamma)$ are considered for the
 three scenarios.
We provide benchmark points with the MSSM spectrum, making the 
models suitable for a detailed phenomenological analysis.

\end{abstract}

\thispagestyle{empty}
\clearpage

\tableofcontents

\section{Introduction}

The minimal supersymmetric
standard model (MSSM) has been subject to a vast amount of study during the past decades. Our ignorance of the
mechanism of supersymmetry (SUSY) breaking as well as the messenger 
mechanism allows
for a large number of free parameters and therefore different
experimental signatures of supersymmetric models (for a review see \cite{hepph0312378}) .

Detailed studies regarding the
concrete low energy spectrum and interactions have to be performed on very specific models. A handful of
benchmark points were chosen \cite{hepph0202233, hepph0106204} in order to
extract low energy implications that can be eventually
contrasted with experiment.
It is however desirable to have a top-down derivation
of soft supersymmetry breaking terms, obtained from a fundamental
theory such as string theory. This would provide a more robust
theoretical motivation for selecting particular benchmark points.

Until recently this task was not fully possible because there were no 
explicit derivations of soft SUSY
breaking terms from string theory (see, for example, \cite{hepph9707209}). Even though some general scenarios were identified, such as dilaton and
moduli domination, there were no proper models in which all moduli were stabilised after SUSY breaking.
Dramatic progress has been achieved in recent years regarding moduli stabilisation in flux
compactifications. In the type IIB context these have been investigated in detail, with the result that a
large class of moduli are stabilised by turning on fluxes of antisymmetric tensor fields. The remaining
moduli, including the  volume of the compact space, can be stabilised by nonperturbative effects, such as in
the KKLT mechanism \cite{hepth0301240} and related generalisations.

Studies of soft SUSY breaking terms from fluxes were performed recently
by several groups \cite{fluxes}.
The phenomenological implications were explored in ref.~\cite{hepph0502151}. However, these studies, 
while fixing all the complex structure moduli and the dilaton, did not stabilise the volume-like moduli. 
In the KKLT scenario the rest of the
moduli are stabilised but
the effect of SUSY breaking by fluxes is washed out by nonperturbative effects. The source of SUSY
breaking is then the least understood part of the scenario corresponding to the introduction of explicit soft
breaking from anti D3 branes.
Furthermore, in this scenario, it has not been possible to perform a proper analysis of soft terms in concrete
models due to technical complications in the stabilisation process.
A phenomenological analysis of models inspired by the KKLT scenario was
recently performed in \cite{hepth0503216, hepph0507110, hepph0504037,
hepph0504036, hepph0511320}.

Fortunately,  a generalisation of the KKLT scenario was constructed
recently \cite{hep0502058},  henceforth referred to as the large
volume scenario, 
in which, for half of the Calabi-Yau
compactifications, all moduli are stabilised. 
The salient feature of these models is that the Calabi-Yau
volume is generically exponentially
large, allowing for a  range of values for the string scale (from TeV
to the GUT scale). This is to be contrasted with the KKLT case in which the volume
has only a logarithmic dependence on flux parameters and it is
difficult to obtain a weakly coupled  (large volume) model.
Another important feature is that only one of the K\" ahler moduli
is stabilised at a large value, whereas all others are\footnote{Unless
  units are explicitly specified, we use string-scale units $m_s=1$.}  
$\gsim1.$
In ref.~\cite{hepth0505076}, 
masses of both bulk and brane moduli were explicitly
computed. The moduli masses are dependent on two parameters - the
Calabi-Yau volume, $\cal{V}$, which is expected to be large, and the
flux superpotential $W_0$ which is generically ${\cal{O}} (1).$ 
Furthermore, unlike the original KKLT version,  the $1/{\cal{V}}$ expansion
allows for explicit control of the calculation of the soft breaking 
terms. These were derived in ref.~\cite{hepth0505076} for the
Calabi-Yau compactifications that admit the exponentially large volume minimum. 

In this article 
we make a study of the phenomenological implications of 
the different scenarios that emerge from this general class of 
string compactifications. We will identify three semi-realistic scenarios
as follows.
\begin{enumerate}
\item{} {\it Generalised Fluxed MSSM.} If the Standard Model lies on D7
  branes, the discrete input parameters allow for string scales
  between the intermediate $10^9$GeV and the GUT scales. The models
  generically suffer from the cosmological moduli problem and smaller
  string scales ($< 10^{11}$GeV) tend to lead to a stop
  lightest supersymmetric particle (LSP) instead of a neutralino. To leading order in
  $1/{\cal{V}}$ expansion, the GUT scale case reproduces the fluxed
  MSSM \cite{hepph0408064}.

\item{}{\it Intermediate Scale Split SUSY.} If the Standard Model lives
  on a set of D3 branes, the scalar masses are naturally many orders of magnitude
  heavier than gaugino masses and for an intermediate string scale
  gaugino masses are of order TeV. Standard fine tuning is then needed
  to keep the Higgs light as in the split supersymmetry
  scenario. Thus we provide a stringy realisation of split SUSY.

\item{}{\it Stringy mSUGRA.} Tuning the value of the flux superpotential $W_0$
  it is possible to obtain a string scale of order the GUT scale and all soft-breaking
  terms of the same order as in the standard mSUGRA scenario. Again,
  this provides a stringy realisation of this popular scenario. 

\end{enumerate}

In each case we perform an RG flow to low energies and impose
phenomenological constraints from dark
 matter, $(g-2)_\mu$ and $BR(b\rightarrow s\gamma)$.
Since the third scenario has been largely explored, we only provide
detailed analysis for models within the first two scenarios, selecting
benchmark points in which the full low-energy spectrum is computed.
In the rest of the article we describe each of these scenarios,
starting with a brief overview of the results in \cite{hepth0505076}.

\section{Moduli Stabilisation, Masses and Scales}
\label{bulksection}
Here we will mention the relevant properties of the KKLT/large volume
compactifications that will be needed for the analysis of soft
breaking terms.
 
Type IIB string models have the following bosonic spectrum of massless fields
in 10d: the metric $g_{MN}$, two rank-two antisymmetric tensors
$B_{MN}, C_{MN}$, a complex dilaton/axion  scalar $S=e^{-\phi} +i a$ and
a rank-four antisymmetric tensor with self-dual field strength. 
A typical string compactification is determined by the following
procedure:

\begin{itemize}

\item
First turn on a background value of the metric to split the spacetime into
our 4d spacetime and six extra compact dimensions which are taken to
be a Calabi-Yau orientifold in order to preserve ${\cal {N}}= 1$
supersymmetry in 4d. 
The Calabi-Yau space is characterised by
the number of two-cycles and their dual four-cycles, as
well as the number of three-cycles. The sizes of these cycles are in
general arbitrary. They define the K\"ahler structure moduli $T$  for the
size of the four and two cycles and complex structure moduli $U$ for the
size of the three cycles. These moduli correspond to the internal
components of the metric which in the 4d effective field theory appear as a set of massless
scalar fields, clearly in contradiction  with experiment. Fixing the
vacuum expectation value of these fields while providing them a mass is the
problem of moduli stabilisation in string theory.

\item
Turning on the other bosonic fields in the spectrum helps stabilise
the complex structure moduli by the standard requirement of flux
quantisation of the field strength of rank-two antisymmetric
tensors. This also fixes the vacuum expectation values of the dilaton/axion field.
Fluxes modify the geometry but in type IIB string theory the remnant
space is a warped Calabi-Yau which is conformally equivalent to a
Calabi-Yau. Therefore the usual properties of Calabi-Yau spaces can still be used
in flux compactifications. This is not the case for the other
string theories, such as the type IIA or heterotic string.

\item
These models generically have D-branes. D3 and
D7 branes can be introduced while preserving supersymmetry. These play an important 
role because it is on the D-branes that the Standard Model can live. Branes
cannot be introduced arbitrarily because of consistency (tadpole
cancellation) conditions requiring the total charge of a given D-brane
type to vanish due to the compactness of the internal manifold.
We can consider the Standard Model matter being on either the D7 or D3
branes. There can also be hidden sector D3/D7 branes. These can induce
non-perturbative superpotentials for the $T_a$ fields ($W_{np}\sim e^{-aT}$), completing
the geometric moduli stabilisation process. 
Other moduli, corresponding to the position of the D7 branes within the
Calabi-Yau manifold, can also  be determined by the combined flux and non-perturbative
superpotentials. 

\item
The previous procedure usually fixes the moduli with a negative value
of the cosmological constant. There are several proposals for lifting
the minimum of the scalar potential to a zero or positive vacuum
energy, such as the inclusion of anti D3 branes \cite{hepth0301240},
D-terms
\cite{bkq},  IASD fluxes \cite{ss}, etc.

\end{itemize}

Only after all moduli have been stabilised is a vacuum obtained
for which the physical properties of the particles (in particular
the soft SUSY breaking terms) in the
spectrum may be analysed. Therefore, phenomenological analysis relies
heavily on moduli stabilisation. In the past, it was simply assumed that moduli were stabilised,
without providing a mechanism.

In the KKLT scenario and recent modifications moduli stabilisation is possible. We
will consider the large volume scenario  described in the introduction since
it allows for explicit minima of the scalar potential at large enough
volumes to trust the effective field theory treatment. 
An important property of this scenario is that, contrary to the
KKLT case, the main source of supersymmetry breaking are the
fluxes and not the {\it ad hoc} introduction of anti D3 branes or D-terms.

The
input parameters are:

\begin{enumerate}
\item{}
The flux superpotential $W_0$. It depends on combinations of integers
determined by the fluxes. Statistically it can take any value but very
small values ($\sim 10^{-11}$) are difficult to obtain. This is usually described
as `fine-tuning' of flux superpotentials.

\item{}
The string coupling constant $g_s$. This corresponds to the vacuum
expectation value (vev) of the
dilaton field $S$ and it is determined by the fluxes. Values of
order $1/10$ are easy to obtain. 

\item{} The volume of the Calabi-Yau $\cal{V}$ which is a combination of the $T$
  moduli.
It is  determined primarily by an exponential dependence on the rank of the hidden sector
  gauge group and the string coupling constant.

\item{} Warp factors of the metric at the location of the D
  branes. These play a role in tuning the minimum to de Sitter
  space. They may also play a role in redshifting scales in the Standard
  Model brane.
\end{enumerate}

The string scale and the gravitino mass are determined in terms of $W_0$ and $\cal{V}$ by
\begin{eqnarray}
m_s &=& \frac{g_s}{\sqrt{4 \pi {\cal{V}}}} M_P,\\
m_{3/2} &=& \frac{g_s^2 W_0}{\sqrt{4 \pi} {\cal{V}}} M_P,
\end{eqnarray}
respectively. Here $g_s$ denotes the string coupling and $M_P= 2.4\times 10^{18}$GeV is the
reduced Planck mass.
Table \ref{d3masstable} shows the dependence of D3 soft terms on $W_0$ and $\cal{V}.$
\begin{table}\renewcommand{\arraystretch}{1.7}
\centering
\vspace{3mm}
\begin{tabular}{|c|c|}
\hline
Quantity & Order of magnitude\\
\hline\hline
\textrm{Scalar masses} $m_i$ & $\frac{g_s^2}{({\cal{V}})^{7/6}}W_0 M_P$\\
\textrm{Gaugino masses} $M_{D3}$ & $\frac{g_s^2}{({\cal{V}})^2}W_0 M_P$\\
\textrm{Scalar trilinear coupling} $A$ & $\frac{g_s^2}{({\cal{V}})^{4/3}}W_0 M_P$\\
$\mu$-\textrm{term} $\hat{\mu}$ & $\frac{g_s^2}{({\cal{V}})^{4/3}}W_0
M_P$\\
\textrm{B term} $\hat{\mu}B$ &
 $\frac{g_s^2}{({\cal{V}})^{7/3}}W_0 M_P$\\
\hline
\end{tabular}
\caption{Soft terms for D3 branes (AMSB contributions not included).\label{d3masstable}}
\end{table}

In the original KKLT model the value of $W_0$ had to be very
small (typically of order $10^{-4}-10^{-11}$) in order to obtain a
minimum within the supergravity approximation. In the scenario of \cite{hep0502058,hepth0505076}, this is
not needed and moduli can be stabilised with the generic case
$W_0\approx 1$. There is, however,
still some freedom to tune $W_0$ in order to explore possible
realistic models.

Table \ref{bulkmodulitable} shows the masses 
of bulk moduli for a range of string scales, as derived in
\cite{hepth0505076}. $W_0$ is taken to be
equal to 1. Note that complex structure and most of the K\" ahler moduli, 
$\tau_s$, have masses comparable to the gravitino mass, while the
modulus with the exponentially large value $\tau_b$ is much lighter.

\begin{table}\renewcommand{\arraystretch}{1.7}
\begin{center}
\begin{tabular}{|c|c|c|}
\hline
String scale $m_s$ & complex structure, $\tau_s$ & $\tau_b$ (lightest bulk modulus)\\
\hline\hline
$10^3$GeV & $1.5\ti 10^{-11}$ & $2.2\ti 10^{-25}$\\
\hline
$10^5$GeV & $1.5\ti 10^{-7}$ & $2.2\ti 10^{-19}$\\
\hline
$10^7$GeV & $1.5\ti 10^{-4}$ & $2.2\ti 10^{-13}$\\
\hline
$10^9$GeV & $15$ & $2.2\ti 10^{-7}$\\
\hline
$10^{11}$GeV & $1.5\ti 10^5$ & $0.22$\\
\hline
$10^{13}$GeV & $1.5\ti 10^9$ & $2.2\ti 10^5$\\
\hline
$10^{15}$GeV & $1.5\ti 10^{13}$ & $2.2\ti 10^{11}$\\
\hline
\end{tabular}
\end{center}
\caption{Bulk moduli masses for a range of string scales assuming
  $W_0\sim 1$. All of the masses are in GeV.
\label{bulkmodulitable}}
\end{table}

There are a few constraints on the scalar moduli masses which restrict
the allowable range of string scales. Firstly, there are fifth force
constraints excluding gravitationally coupled scalars
lighter than about $10^{-4}$eV. There is also the cosmological moduli
problem \cite{hepph9308292, hepph9308325, ross}
which states that the universe might be overclosed if moduli
masses are in the range $10^{-7} {\rm GeV}<m<10^4 {\rm GeV}$.
The lightest modulus in the scenario is $\tau_b$ whose mass behaves
as ${\cal{V}}^{-3/2}.$ Setting its mass equal to the lowest allowed
by fifth force constraints, we get a lower bound on the string scale
of about $10^8$GeV. However, for this case the masses of the other bulk
moduli are such that the cosmological moduli problem might be
relevant. The string scale for which all bulk moduli are heavier than
$10^4$ GeV is around $m_s = 7.7\times 10^{12}$GeV.

Table \ref{branemodulitable} shows the values of D3 brane soft
breaking terms and the gravitino mass
for a range of allowed string scales $m_s.$ 
Results for D7 soft parameters are not shown since they are all
of the same order as the gravitino mass. The estimates of anomaly mediated
SUSY breaking (AMSB) contributions are obtained
as the gravitino mass multiplied by a loop suppression factor. This will
later be shown to be too naive an estimate, at least for gaugino masses.
\begin{table}\small{\renewcommand{\arraystretch}{1.7}
\begin{tabular}{|c|c|c|c|c|c|c|}
\hline
$m_s$ & $m_{3/2}$ & D3  & D3 & D3  & AMSB  & AMSB \\
GeV & GeV & scalars & gauginos & A-terms & scalar & gaugino\\
\hline\hline
$10^{10}$  & 1500 & 7.8 & $3.2\ti 10^{-11}$ & 0.041 & 21 & 19\\
\hline
$10^{11}$  & $150\ti 10^3$ & $1.7\ti 10^3$ & $3.2\ti 10^{-7}$ & 19 & 
$2.1\ti 10^3$ & $1.9\ti 10^3$\\
\hline
$10^{12}$  & $1.5\ti 10^7$ & $3.6\ti 10^5$  & 0.003 & $8.9\ti 10^3$ &
$2.1\ti 10^5$ & $1.9\ti 10^5$ \\
\hline
$10^{13}$ & $1.5\ti 10^9$ & $7.8\ti 10^7$ & 32 & $4\ti 10^6$
& $2.1 \ti 10^7$ & $1.9\ti 10^7$\\
\hline
$10^{14}$ & $1.5\ti 10^{11}$ & $1.7\ti 10^{10}$ & $3.2\ti
10^5$ & $1.9\ti 10^9$ & $2.1\ti 10^9$ & $1.9\ti 10^9$ \\
\hline
$10^{15}$ & $1.5\ti 10^{13}$ & $3.6\ti 10^{12}$ & $3.2\ti
10^9$ & $8.8\ti 10^{11}$ & $2.1\ti 10^{11}$ & $1.9\ti 10^{11}$\\
\hline
$10^{16}$ & $1.5 \ti 10^{15}$ & $7.8\ti 10^{14}$ & $3.2\ti
10^{13}$ & $4.1\ti 10^{14}$ & $2.1\ti 10^{13}$ & $1.9\ti 10^{13}$\\
\hline
\end{tabular}
\caption{Orders of magnitude of D3 brane soft breaking terms and the gravitino mass for a range
of string scales. All the results are in GeV.\label{branemodulitable}}}
\end{table}

Having listed the generic values for masses of various moduli for a range
of values of the string scale, we can pick out a few exemplary scenarios and
analyse their phenomenology in more detail.
We restrict ourselves to considering all MSSM matter being put exclusively 
on D3 or D7 branes, without D3-D7 fields. 
A particular compactification yielding the MSSM spectrum in this
sub-class of models is yet to be found. In this paper we take
the bottom up approach and assume that this has been achieved.
Some possible constructions, albeit not fixing all the moduli, are
described in \cite{hepth0005067, hepth0312051, hepph0105042,
hepth0508089, hepth0503079}.

\section{Matter on D7 branes}

\subsection{Generalised Fluxed MSSM}

 If matter is put on D7 branes only, the large volume
 compactifications give rise to leading order in the $1/\cal{V}$
 expansion to the fluxed MSSM soft terms derived in 
\cite{fluxes}. The gaugino
masses, A-terms, scalar masses, and B-term are determined in terms of the 
gravitino mass $m_{3/2}=M$ by
\begin{equation}
m_{1/2} = M, A = -3 M, m_0 = |M|, B=-2 M,
\end{equation}
where the $B$ term in the scalar potential is taken to be
$-m_3^2 H_1 H_2 = -\mu B H_1 H_2.$
It was shown in \cite{hepth0505076} that the inclusion of nonperturbative and
$\alpha'$ effects does not affect this computation significantly.

The phenomenology of this type of models was investigated in 
ref.~\cite{hepph0502151}, where it was assumed that the couplings unify at the
GUT scale $M_{GUT}\sim 10^{16}$GeV.
In contrast to this,  we expect the gauge couplings to unify at the string
scale, which may be different from $M_{GUT}$, as shown in Table \ref{branemodulitable}.

The large volume models impose a relationship between the
gravitino mass $M$ and the string scale $m_s.$ 
The string scale required for obtaining TeV size scalar masses, without any 
fine tuning in $W_0$, is of the order
$10^9-10^{10}$GeV. However, it is well known that with the MSSM spectrum alone,
the gauge couplings unify at approximately $10^{16}$GeV.
In order to obtain gauge unification at a lowered scale, 
extra matter at a certain scale must exist to modify the running of the
gauge couplings \footnote{
For certain types of branes at singularities, gauge unification
is not necessary. Similarly this is true of models with intersecting
or magnetised D-branes. We will investigate this possibility later in the text.}. We choose this scale to be 
1TeV in order to avoid introducing a new hierarchy. Since the
vector-like representation of the additional matter allows for explicit mass terms, 
they are expected to be of the order of the string scale and must therefore be lowered
to $1$TeV by some mechanism. We will also assume that the Yukawa couplings
of the extra matter to MSSM matter are negligible.
To find out what extra matter needs to be added in order to achieve
gauge unification, we may use the 1-loop
RGEs for the MSSM including extra matter, found in the Appendix of
\cite{hepph9703293}. The relevant equations are
\begin{eqnarray}
16\pi^2 {dg_1\over{dt}} &=& g_1^3 \left( {33\over5} + {n_Q\over5} +
{8n_U\over5} + {2n_D\over5} + 
{n_S\over10} +{3\over5} n_2 + {6\over5}n_E\right)\\
16\pi^2 {dg_2\over{dt}} &=& g_2^3 \left( 1+ 3n_Q+n_2\right)\\
16\pi^2 {dg_3\over{dt}} &=& g_3^3 \left( -3 + 2n_Q + n_U + n_D + n_S \right).
\end{eqnarray}
Here $g_1, g_2, g_3$ are the gauge couplings for $U(1), SU(2)$ and
$SU(3)$, respectively ($g_1$ being GUT normalised), while $t=\ln\mu$
with $\mu$ the $\overline{DR}$ renormalisation scale.
$n_2$ is the number of additional vector-like lepton superfield doublets $L + \bar{L}$
and $n_E$ the number of right handed lepton singlets
$E + \bar{E}.$ $n_Q, n_U, n_D, n_S$
are the numbers of $Q+\bar{Q}, U+\bar{U}, D+\bar{D}$ and exotic ``sextons''
$S$ which are colour triplets, electroweak singlets and have $Y=1/6.$
It is straightforward to see that lowering the string scale is easily
achieved by using nonzero values only for $n_2, n_E.$
A one loop analysis shows that the choice $n_2 = 4, n_E = 6$ unifies the 
couplings at approximately $1.2\times 10^9$ GeV.
To study the renormalisation group behaviour, we
use \verb+SoftSUSY1.9+ \cite{hepph0104145} modified so that RGEs with extra matter are used
above the scale of $1$TeV. A 1-loop analysis
is performed, giving a sufficient level of accuracy for our needs.
The values used for Standard Model input parameters are the current central
values: $m_t = 172.7$GeV
\cite{hepex0507091}, $m_b (m_b)^{\overline{MS}} =4.25$GeV, ${\alpha}_s (M_Z) = 0.1187$,
$\alpha^{-1} (M_Z)^{\overline{MS}} = 127.918$, $M_Z = 91.1187$GeV \cite{pdg}.

\begin{figure}
\includegraphics{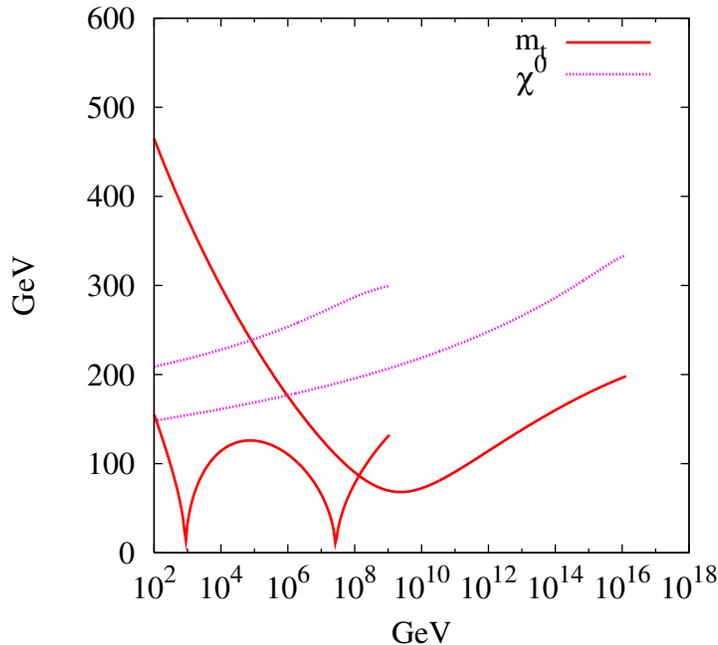}
\caption{Evolution of the lightest stop and neutralino masses
for the $m_s\sim 10^9$GeV and $m_s = m_{GUT}$ scenarios.
The tree level stop mass squared is negative for renormalisation mass scales in the
range $10^3-10^7$GeV in
the former case, so $\sqrt{|m_{\tilde{t_1}}^2|}$ is plotted.
The boundary condition is $M=350$GeV and $\tan \beta = 10$, $\mu$
positive. \label{softmassevofig}}
\end{figure}

For $m_s \sim 10^9$GeV, the RGE analysis results in the stop being the LSP. This did 
not occur in the scenario with GUT scale unification. The reason is that
the scalar masses do not evolve sufficiently for a lowered string
scale. 
The mass of the lightest stop $\tilde{t}_1$ is determined by the diagonalising the matrix
\begin{eqnarray}
\left( \begin{array}{cc}
(m^2_{\tilde{L}})_{33}+m^2_{t} + (1/2 - 2s_W^2/3)M_Z^2 c_{2\beta}  & 
   m_{t} ((A_U)_{33} - \mu \cot \beta) \\
 m_{t} ((A_U)_{33} - \mu \cot \beta) & (m^2_{\tilde{u}R})_{33}+m^2_{t}
+ (2/3) s_W^2 M_Z^2 c_{2\beta}
\end{array} \right).
\end{eqnarray}
Here $c_{2\beta} = \cos{2\beta}$ and $s_W$ is the sine of the weak
mixing angle. 
In Figure \ref{softmassevofig} the evolution of the
tree level stop mass is shown together with the
lightest neutralino mass, for both the intermediate scale and GUT
cases. Even though the tree level mass squared becomes negative at
intermediate renormalisation scales, we do not anticipate a charge
and colour breaking minimum \cite{hepph9507294}. This is because
the physical mass squared (a better approximation to the
relevant term in the effective potential than the tree-level mass)
is positive.

Since the MSSM spectrum of these models does not provide a natural
candidate for dark matter \cite{heavyisotope}, we have to look for one 
elsewhere, e.g. in the moduli sector. Also, the models will have to
be R-parity violating so that the light stop decays quickly and does
not spoil the successful predictions of nucleosynthesis.
Another potential difficulty in the cosmological context could be
posed by the light bulk moduli for the string scale of order $10^9$GeV,
as pointed out in Section \ref{bulksection}.

The models can still be subjected to other
experimental constraints, such as the ones coming from precision 
measurements of the $b\to s\gamma$ branching ratio and the anomalous 
magnetic moment of the muon.

The most recent average measurement of the $b\to s\gamma$ branching 
ratio may be obtained from \cite{heavyflavour} and is
$(3.39\pm 0.30) \ti 10^{-4}.$ The theoretical
uncertainty in this result is \cite{hepph0410155} $0.30\ti 10^{-4}$;
adding the two errors in quadrature, we obtain the $1\sigma$ bound,
$BR(b\to s\gamma) = (3.4\pm 0.4)\times 10^{-4}.$

We also impose limits on the new physics contribution to $a_\mu =
(g_\mu-2)/2.$ The experimental value of $a_\mu$ is
$(11659208\pm6)\ti 10^{-10}$ \cite{hepex0401008}. The Standard Model 
computation yields \cite{hepph0411168,hepph0402285} 
$(11659189\pm 6)\ti 10^{-10}$ for the same quantity. 
This results
in the $1\sigma$ bound on the non-SM contribution to $a_\mu$,
$\delta a_{\mu} = (19\pm 9)\times 10^{10}.$
It is important to note that $\delta a_{\mu}$ usually has the
same sign as $\mu$, so $\mu>0$ is preferred by experiment.

We also impose bounds on the Higgs mass obtained by the LEP2
collaborations \cite{hepex0306033}. The lower bound is $114.4$ GeV
at the 95\% CL. The error in theoretical predictions is estimated to
$3$GeV so we require $m_h>111$GeV on the \verb+SoftSUSY1.9+ prediction.

One can consider a simple modification of this scenario with a raised
string scale, by allowing some fine tuning in $W_0$. Namely, one can
decrease ${\cal{V}}$ while decreasing $W_0$, hence increasing $m_s$
but also decreasing the amount of SUSY breaking.
Let us consider the cases $m_s\sim 10^{12}$GeV and
$m_s\sim 10^{14}$GeV\, \footnote{It should be noted that lifting the string scale does
not remove the cosmological problems associated with light bulk moduli, since masses of some of 
these are proportional to $W_0.$}. In the former case, to make the gauge 
couplings unify at the string scale, we add $n_2 = 2, n_E = 3.$
In the $m_s\sim 10^{14}$GeV case, we need to add $n_2 = 1, n_E=1.$
Interestingly, already in the $m_s\sim 10^{12}$GeV case, the LSP
becomes a neutralino.
We can then apply the hypothesis that the neutralino constitutes all
of the cold dark matter relic density.
The WMAP \cite{astroph0302209,astroph0302207} constraint on the relic 
density of dark matter particles (at the $3\sigma$ level) is 
\begin{equation}
0.084 < \Omega h^2 < 0.138.
\end{equation}

However, the $B=-2M$ condition cannot be satisfied for any values of
$m_s, M$ with $\mu>0$
and can only be satisfied for $\mu<0$ for string scales $10^{14}$GeV
and above. To see this, one notes that the EWSB conditions
\begin{eqnarray}
\mu^2 &=& {{-m_{H_1}^2 \tan^2 \beta + m_{H_1}^2}\over{\tan^2\beta-1}} -
          {1\over2} M_Z^2,\\
\mu B &=& {1\over2} \sin 2\beta (m_{H_1}^2 + m_{H_2}^2 + 2\mu^2)
\end{eqnarray}
determine the values of $B$ and $\mu$ at the scale 
$M_{SUSY}=\sqrt{m_{\tilde{t}_1} m_{\tilde{t}_2}}$
in terms of $\tan\beta$ and the soft breaking masses $m_{H_1}, m_{H_2}.$ 
One then needs to evolve back to the string scale to check whether
the condition $B=-2M$ can be satisfied.

\begin{figure}
{\twographs{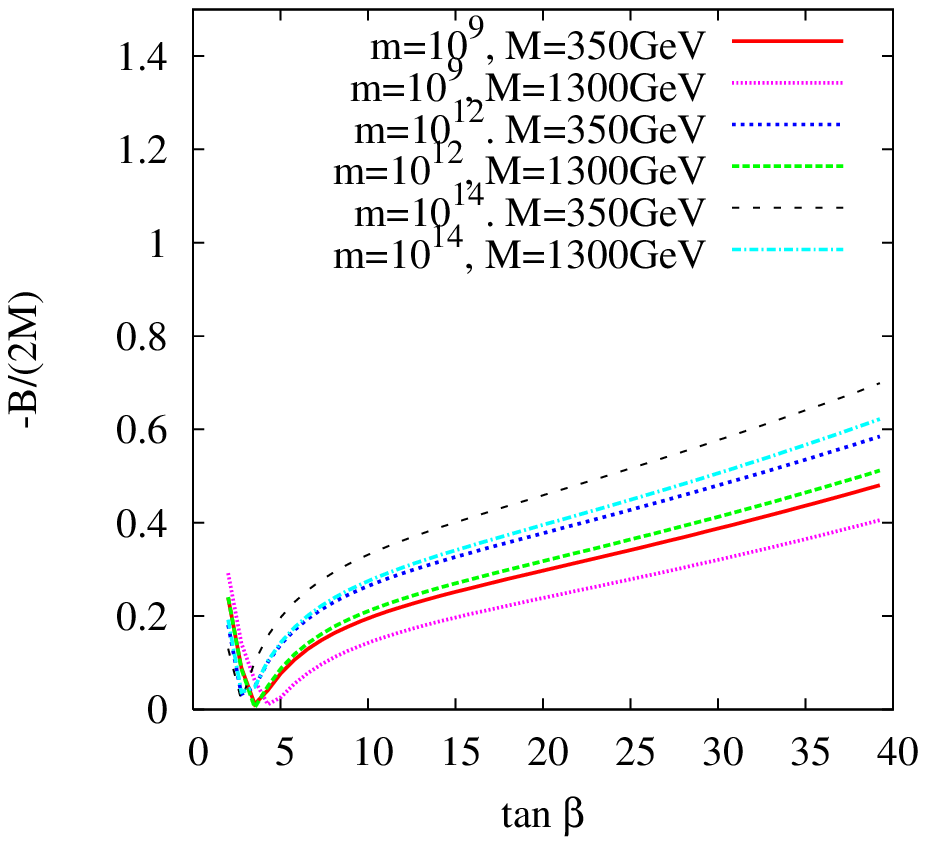}{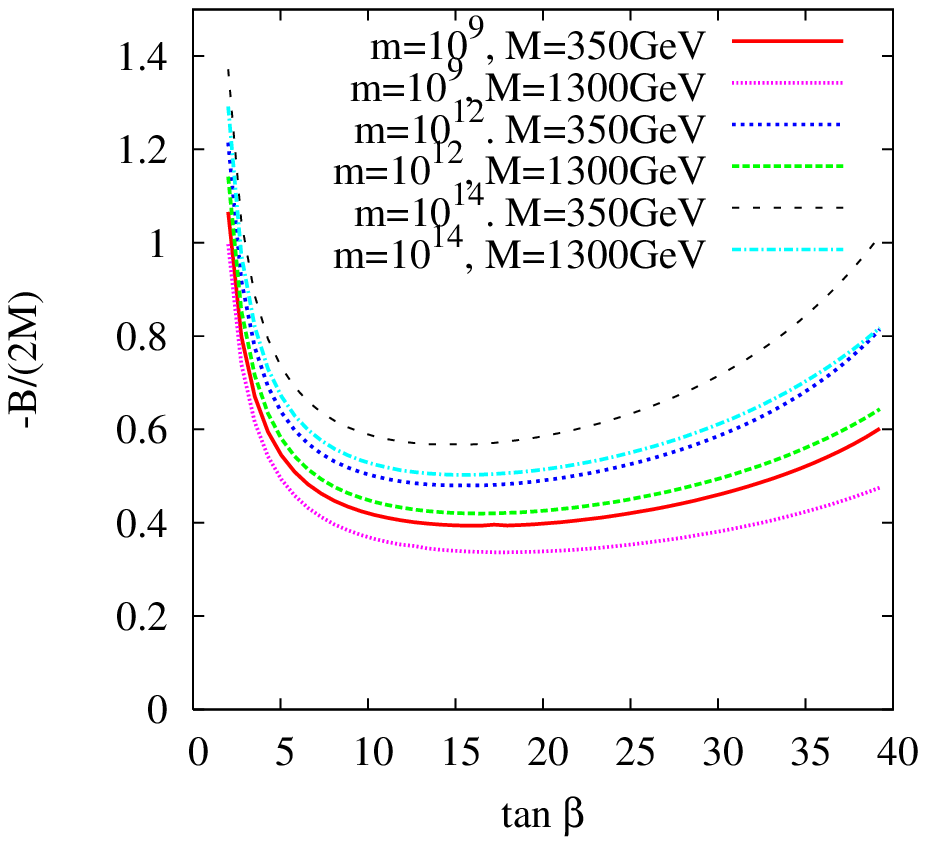}}
\caption{Ratio $B({m_s})/2M$ for (a) $\mu>0$ and (b) $\mu<0$, $M=350$GeV
and $M=1300$GeV.\label{btomratiofig}}
\end{figure}

The dependence of the ratio $|B|/(2M)$ on $\tan\beta$ is displayed in 
Figure \ref{btomratiofig}, for various choices of string
scale. Solutions to $B=-2M$ exist only for the low $\tan\beta$ region 
and $\mu<0.$ However, it can be verified that in 
this region, the Higgs is too light compared to the LEP2 bound.

Therefore we will neglect for the time being the boundary condition
$B=-2M$, assuming that the values for the $\mu$ and $B$-terms required
for correct electroweak symmetry breaking (EWSB) are generated by some
unknown mechanism.

\begin{figure}
{\twographs{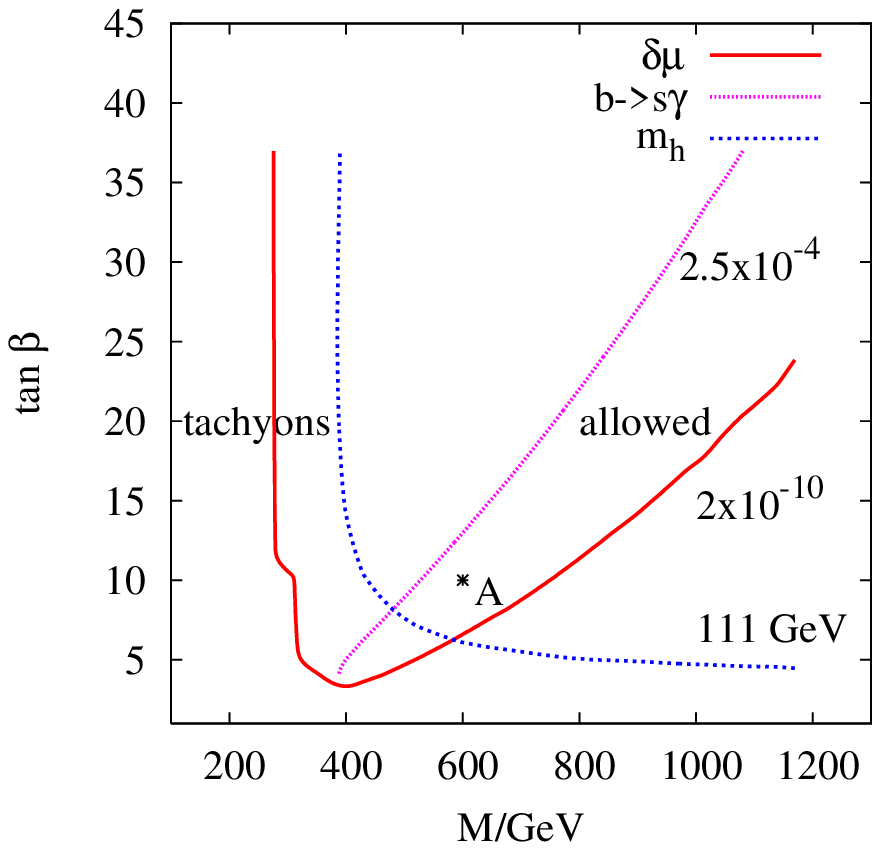}{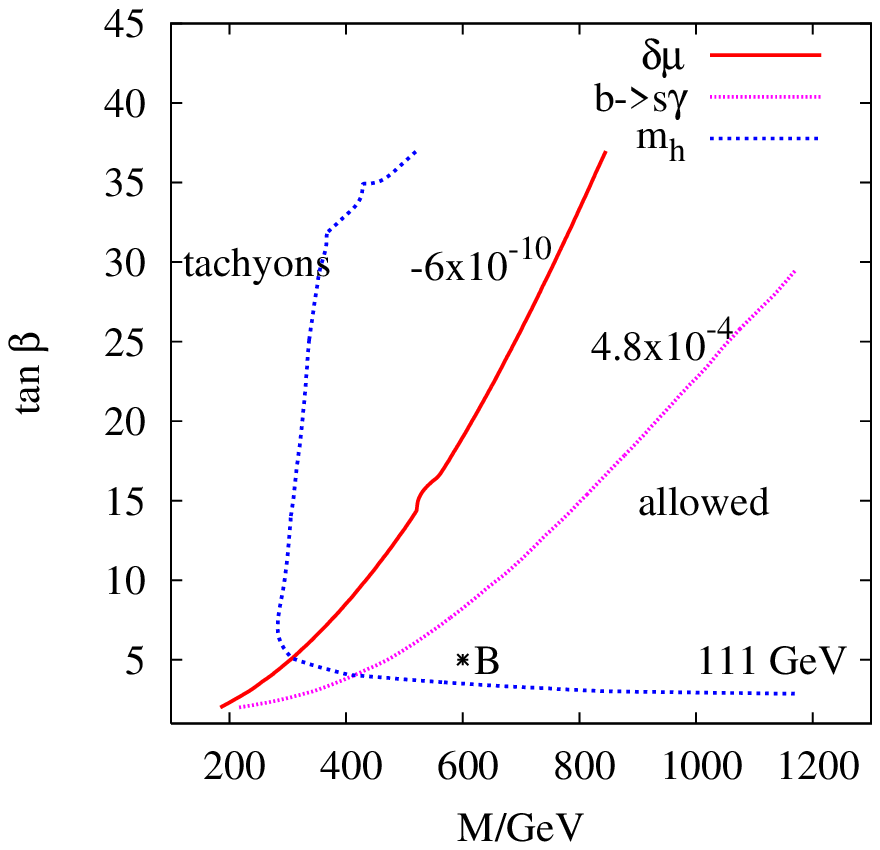}}
\caption{Contour plots of $\delta a_{\mu}$ and $BR(b\to s\gamma)$ on 
$\tan\beta$ and $M$ for (a) $\mu>0$ and (b) $\mu<0$ and $m_s\sim 10^9$ GeV.
$2\sigma$ bounds are used for $\mu>0$ while $3\sigma$ ones are used
for $\mu<0.$
\label{bsgamma10to9}}
\end{figure}

\begin{figure}
{\twographs{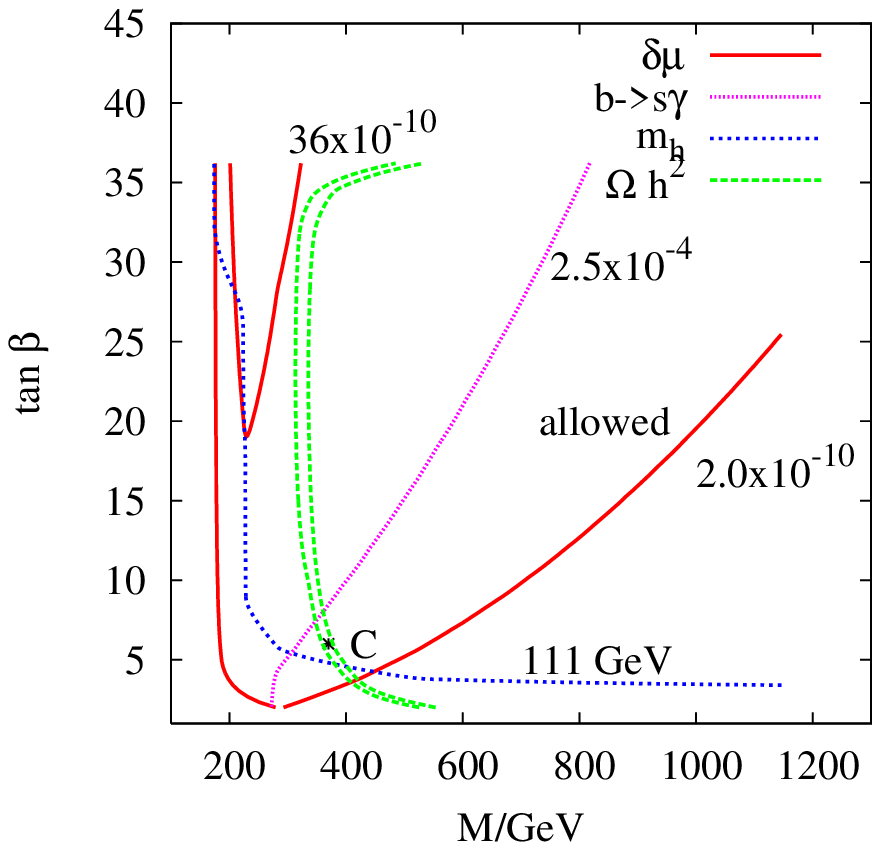}{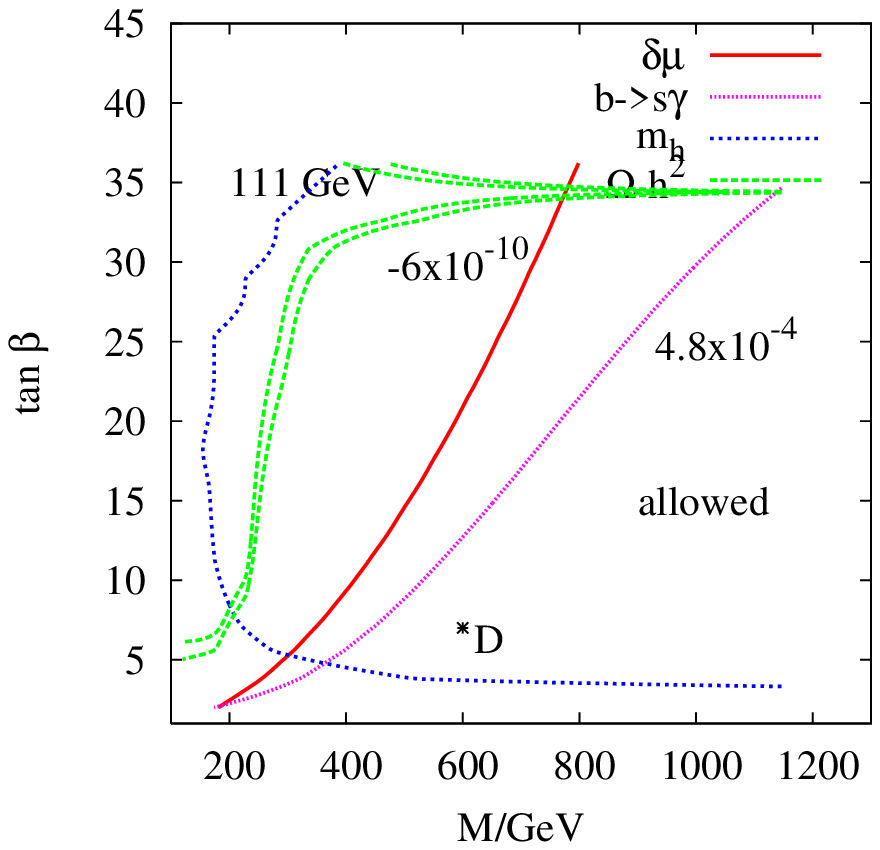}}
\caption{Contour plots of $\delta a_{\mu}$, $BR(b\to s\gamma)$ and
$\Omega h^2$ on
$\tan\beta$ and $M$ for (a) $\mu>0$ and (b) $\mu<0$ and $m_s\sim 10^{12}$ GeV.
$2\sigma$ bounds are used for $\mu>0$ while $3\sigma$ ones are used
for $\mu<0.$
\label{bsgamma10to12}}
\end{figure}

There are then two free parameters, the mass scale $M$ and $\tan \beta.$ 
The $b\to s\gamma$ branching ratio, $\delta a_{\mu}$ and $\Omega h^2$ are computed using the
\verb+micrOmegas1.3+ package \cite{hepph0405253} interfaced with 
\verb+SoftSUSY1.9+ via the SUSY Les Houches Accord \cite{hepph0311123}. The results are plotted in 
Figure \ref{bsgamma10to9} for $m_s\sim 10^9$GeV and in Figure
\ref{bsgamma10to12} for $m_s\sim 10^{12}$GeV.
Although the dark matter relic density computation is sensitive to the details of the 
low energy spectrum, the allowed region in the $M-\tan\beta$ plane should not shift significantly when 
two-loop RGE equations are used instead of one-loop ones.
Note that for $\tan\beta$ greater than around 40 one cannot obtain
correct EWSB and those values are not included in the graphs.
Because the LSP is the stop for $m_s\sim 10^9$GeV, we do not plot $\Omega h^2$ in Figure
\ref{bsgamma10to9}.

MSSM spectra for a few sample points in the parameter space for which
both the $\delta a_\mu$ and $BR(b\to s\gamma)$ constraints are satisfied
are shown in Table \ref{spectratable}. The points chosen are denoted
by A,B in Figure \ref{bsgamma10to9} and C,D in Figure 
\ref{bsgamma10to12}. Point C satisfies all the constraints, including
the $\chi_1^0$ cold dark matter hypothesis. The other points satisfy
the $BR(b\to s\gamma)$ and $(g-2)_{\mu}$ constraints. In the $m_s\sim 10^{12}$GeV,
$\mu<0$ case, it is not possible to satisfy the dark matter constraint.

\begin{table}
\begin{center}
\begin{tabular}{|c|c|c|c|c|}
\hline
& A & B & C & D \\
\hline
$m_s$ & $10^9$ & $10^9$ & $10^{12}$ & $10^{12}$\\
\hline
$\tan\beta$ & 10 & 5 & 6 & 7\\
\hline
$M$ & 600 & 600 & 370 & 600\\
\hline
${\rm sgn} \mu$ & + & - & + & -\\
\hline
\hline
$\tilde{e}_L, \tilde{\mu}_L$ & 685 & 684 & 436 & 702 \\
\hline
$\tilde{e}_R, \tilde{\mu}_R$ & 633 & 633 & 395 & 637 \\
\hline
$\tilde{\tau}_L$ & 681 & 684 & 437 & 699  \\
\hline
$\tilde{\tau}_R$ & 618 & 630 & 389 & 630  \\
\hline
$\tilde{u}_1, \tilde{c}_1$ & 973 & 973 & 701 & 1099  \\
\hline
$\tilde{u}_2, \tilde{c}_2$ & 1013 & 1013 & 728 & 1143  \\
\hline
$\tilde{t}_1$ & 343 & 385 & 237 & 500 \\
\hline
$\tilde{t}_2$ & 885 & 873 & 678 & 973 \\
\hline
$\tilde{d}_1, \tilde{s}_1$ & 968 & 968 & 698 & 1093 \\
\hline
$\tilde{d}_2, \tilde{s}_2$ & 1016 & 1016 & 732 & 1146 \\
\hline
$\tilde{b}_1$ & 807 & 814 & 593 & 924  \\
\hline
$\tilde{b}_2$ & 949 & 958 & 690 & 1078\\
\hline
$\chi_1^0$ & 378 & 381 & 201 & 336\\
\hline
$\chi_2^0$ & 543 & 556 & 321 & 544\\
\hline
$\chi_3^0$ & 764 & 802 & 585 & 916\\
\hline
$\chi_4^0$ & 782 & 807 & 601 & 919\\
\hline
$\chi_1^{\pm}$ & 543 & 556 & 321 & 544\\
\hline
$\chi_2^{\pm}$ & 781 & 808 & 600 & 921\\
\hline
$A_0, H_0$ & 1008 & 1065 & 732 & 1151\\
\hline
$H^{\pm}$ & 1012 & 1067 & 736 & 1153\\
\hline
$\tilde{g}$ & 1014 & 1014 & 736 & 1150\\
\hline
$\tilde{\nu}_{1,2}$ & 680 & 680 & 429 & 698\\
\hline
$\tilde{\nu}_3$ & 674 & 679 & 428 & 694\\
\hline
\hline
$B(b\to s\gamma)/10^{-4}$  &  $2.8$ & $4.4$ & $2.9$ & $4.3$\\
\hline
$\delta a_{\mu}/10^{-10}$ &   $3.0$ &  $-1.6$ & $4.2$ & $-2.0$\\
\hline
$\Omega h^2$ & --- & --- & 0.111 & 2.01\\
\hline
\end{tabular}
\end{center}
\caption{
Sparticle spectra for the intermediate scale models. All masses are in GeV.
\label{spectratable}}
\end{table}

As mentioned before, it is also possible to construct models where gauge unfication
does not take place at the string scale. We investigate that possibility
for completeness, using the standard MSSM RGEs without any extra matter.
In this case, two loop RGEs are used rather than one loop RGEs. The results, however,
turn out not to differ significantly from the unification at $m_s$ scenario.


\section{Matter on D3 branes}

Let us now investigate the semi-realistic scenarios that can be constructed
assuming the Standard Model lives on a set of D3 branes. For this we have to
consider the spectrum in Table \ref{branemodulitable}. Notice that there is a hierarchy
between the scalar and the gaugino masses. The difference between the
two masses decreases when the string scale is increased.  At the GUT
scale they tend to be of the same size but apparently too heavy. For
any other scale the scalars are  much heavier than the gauginos. At an
intermediate string scale the scalar masses are of order 1 TeV but the
gauginos are too light. At first sight we might conclude that anomaly
mediation contribution to the gaugino masses would be dominant but the
no-scale nature of our models is such that this contribution is also
negligible. 

Loop corrections are not large enough to lift the
gaugino masses to realistic values. Only for $A$ terms of order $\sim
10^7$GeV loop corrections induce gaugino masses of order TeV but at
that scale scalar masses are very heavy ($\sim 10^7$GeV). We then
would have to argue that fine-tuning similar to split supersymmetry is
at work to keep the Higgs light.

A second alternative is to consider the case in which the scalar and
gaugino masses are of the same order, which in Table \ref{branemodulitable} 
corresponds to the GUT scale. Again, the value of $W_0$ can
be fine tuned to lower the effective masses of scalars and gauginos to
the TeV range. We will consider next each of these two scenarios.

\subsection{Intermediate Scale Split SUSY}
We choose the string scale to be $\sim 10^{13}$GeV, with the scalar masses
of order $10^7$GeV. One might worry that the gauginos will be extremely heavy 
as well due to the AMSB contribution. However, it was shown in 
\cite{hepth9911029}
that in no-scale type models the AMSB contribution to the gaugino mass
is vanishing (see the Appendix).

There are two kinds of corrections to this result in our models - firstly, we
include perturbative $\alpha'$ corrections in the K\" ahler potential, 
and secondly, we include nonperturbative contributions to the superpotential 
$W.$ At tree level, the gaugino masses are set by the size of $F^S$, which
is induced by nonzero mixing between the dilaton and K\" ahler moduli once
$\alpha'$ corrections are included. However, $F^S$
is proportional to $1/{\cal{V}}^2$ and is very small for large volume. As seen above, the
one loop anomaly-induced contribution also vanishes in the no-scale 
approximation.

In the two K\" ahler modulus model investigated in \cite{hepth0505076}, we have 
$\partial_s K \sim {1\over{\cal{V}}}$ and $\partial_b K \sim
{1\over{\cal{V}}^{2/3}}$ with $s,b$ denoting the small and large K\"
ahler moduli, respectively. The nonperturbative
contributions to the F-terms are $F_s^{np} \sim {1\over{\cal{V}}}$ and
$F_b^{np} \sim {1\over{\cal{V}}^{4/3}}.$ Therefore, the contributions
$K_i F^i_{np}$ to the gaugino mass in formula (\ref{gaugmass}) in the Appendix
scale like $1/{\cal{V}}^2.$ Similarly, the contribution
to gaugino masses from $F^S$ is of the same form. Therefore the anomaly 
contribution is not
as large as would follow from the naive estimate based on just including the
superconformal anomaly contribution proportional to $m_{3/2}.$

Hence we arrive at a scenario with the string scale at around 
$10^{13}$GeV, the scalar
masses at $\tilde{m}\sim 10^{7}$GeV and gaugino masses approximately
vanishing at the string scale. 
The first question to be considered is whether
it is possible to generate gaugino masses sufficiently large to pass
experimental lower bounds through 
RG evolution to lower scales. To answer this, one must consider the
two-loop MSSM RGEs for gaugino masses (as found in \cite{hepph9311340}):
\begin{eqnarray}
\lefteqn{{d\over{dt}} M_a = {2g_a^2\over{16\pi^2}} b_a M_a +} \\
& &{2g_a^2\over{(16\pi^2)^2}} \left( \sum_b B_{ab} g_b^2 (M_a+M_b) +
\sum_{\alpha=u,d,e} C^\alpha_a ({\rm tr} (Y^{\alpha\dag} h^\alpha) - M_a {\rm tr} (Y^{\alpha^\dag} Y^\alpha))
\right)\nonumber.
\end{eqnarray}
The SUSY breaking trilinear scalar couplings  $h^\alpha$ (defined in
terms of the more familiar $A$ notation via
$h^\alpha_{ij} = A^\alpha_{ij} Y^\alpha_{ij}$, $Y^\alpha$ being the
Yukawa couplings)
enter the two loop renormalisation; if they are large
enough it might be possible to generate sufficient gaugino masses in the
running between the string scale and the scalar mass scale.

We will have to assume that the Higgs mass squared is fine-tuned
at the scalar mass scale to be negative and of order $-m_{ew}^2.$
Given this assumption, below scale $\tilde{m}$, the effective field
theory spectrum is that of split
supersymmetry \cite{hepth0405159, hepph0406088}: all the SM particles including one light Higgs doublet $H$,
together with gauginos $\tilde{B}, \tilde{W}, \tilde{g}$ and higgsinos
$\tilde{H}_u, \tilde{H}_d.$ The Lagrangian consists of kinetic terms and
\begin{eqnarray}
\label{splitsusyl}
m^2 H^\dag H - {\lambda\over2} (H^\dag H)^2 - \Bigl( Y_{ij}^u \bar{q}_j u_i
\epsilon H^* + Y_{ij}^d \bar{q}_j d_i H + Y_{ij}^e \bar{l}_j e_i H \nonumber\\
+ {M_3\over2} \tilde{g}^A \tilde{g}^A + {M_2\over2} \tilde{W}^a \tilde{W}^a
+ {M_1\over2} \tilde{B} \tilde{B} + \mu \tilde{H}_u^T \epsilon
\tilde{H}_d
\nonumber\\
+ {H^\dag\over\sqrt{2}} (\tilde{g}_u \sigma^a \tilde{W}^a + \tilde{g}'_u 
\tilde{B} ) \tilde{H}_u +  { {H^T \epsilon}\over\sqrt2} (-\tilde{g}_d
\sigma^a \tilde{W}^a + \tilde{g}'_d \tilde{B}) \tilde{H}_d + {\rm h.c.}
\Bigr),
\end{eqnarray}
where $\sigma^a$ are the Pauli matrices and $\epsilon = i\sigma^2.$
Below the scale of the scale of the scalar masses, we use the RGEs of
split supersymmetry, listed in the Appendix of \cite{hepph0406088}.

The simplest experimental constraint to impose is the one on the Higgs
mass. The Higgs quartic coupling $\lambda$ is matched at the scalar 
mass scale by the formula
\begin{equation}
\label{lambdabc}
\lambda (\tilde{m}) = {{{g^2 (\tilde{m}) + g'^2 (\tilde{m})}}\over4}
\cos^2{2\beta}.
\end{equation}
$g$ and $g'$ are the values of the $SU(2)$ and $U(1)_Y$ gauge couplings,
and the relation between the GUT normalised $g_1$ and $g'$ is
$g_1 = \sqrt{5/3} g'.$
Equation (\ref{lambdabc}) can be obtained by matching the split 
SUSY Lagrangian (\ref{splitsusyl}) with the usual SUSY Lagrangian 
valid above the scalar mass $\tilde{m}.$ It receives finite threshold
corrections of order $A^2/\tilde{m}^2$, but this is negligible in our
models. The boundary conditions for $\tilde{g}_{u,d},
\tilde{g}'_{u,d}$ are similarly given at scale
$\tilde{m}$ in terms of $g, g'$ and $\tan\beta.$
The values of $g, g'$ at the scale $\tilde{m}$ are
found by a simple one-loop evolution from their experimentally
determined values at $M_Z$; after that one evolves
$\lambda$ down to $M_Z$ using the split SUSY RGEs.

\begin{figure}
\includegraphics{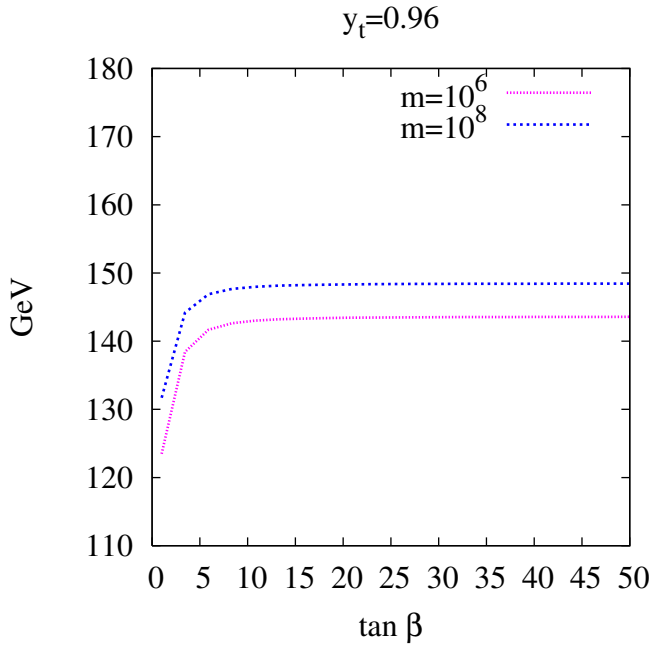}
\caption{The value of the Higgs mass against $\tan\beta$ for $\tilde{m} = 10^6$GeV and
$\tilde{m} = 10^8$GeV.\label{higgstanbfig}}
\end{figure}

The renormalisation effects on $\lambda$ are large and it is natural
to obtain a Higgs significantly heavier than the LEP2 bound.
The Higgs mass is estimated as $m_H = \sqrt{\lambda} v$ with $v=246.22$GeV.
$m_H$ ranges from $\sim 142$ GeV to $\sim 163$ GeV depending
on the values of $\tan{\beta}$, the top mass and the scalar masses
$\tilde{m}.$ 
The principal dependency is through the value of the top Yukawa, i.e.
the top mass. The tree level formula for the Yukawa coupling is
$m_t^{pole}\sqrt{2}= y_t (m_t) v (m_t)$, giving $y_t = 0.99$ for 
$m_t = 172.7$ GeV. The dominant one loop corrections to this are
\cite{hepph9609331}
\begin{equation}
m_t^{pole} \sqrt{2}= y_t (m_t) v (m_t) \left( 1+ {g_3^2 (m_t)
\over{3\pi^2}} - {y_t^2 (m_t)\over{8\pi^2}}
\right),
\end{equation}
giving $y_t (m_t) = 0.96.$
We use the 1-loop corrected value for greater accuracy. As can be seen in Figure 
\ref{higgstanbfig}, the Higgs mass is essentially independent of $\tan\beta$ for large 
values of $\tan \beta.$

\begin{figure}
{\threegraphs{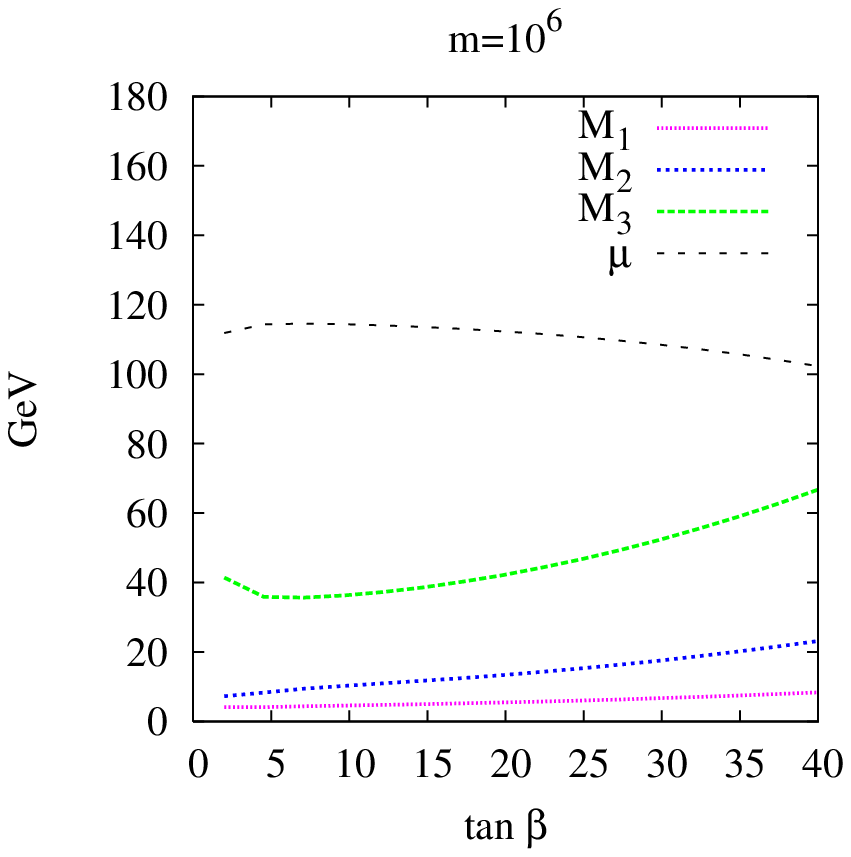}{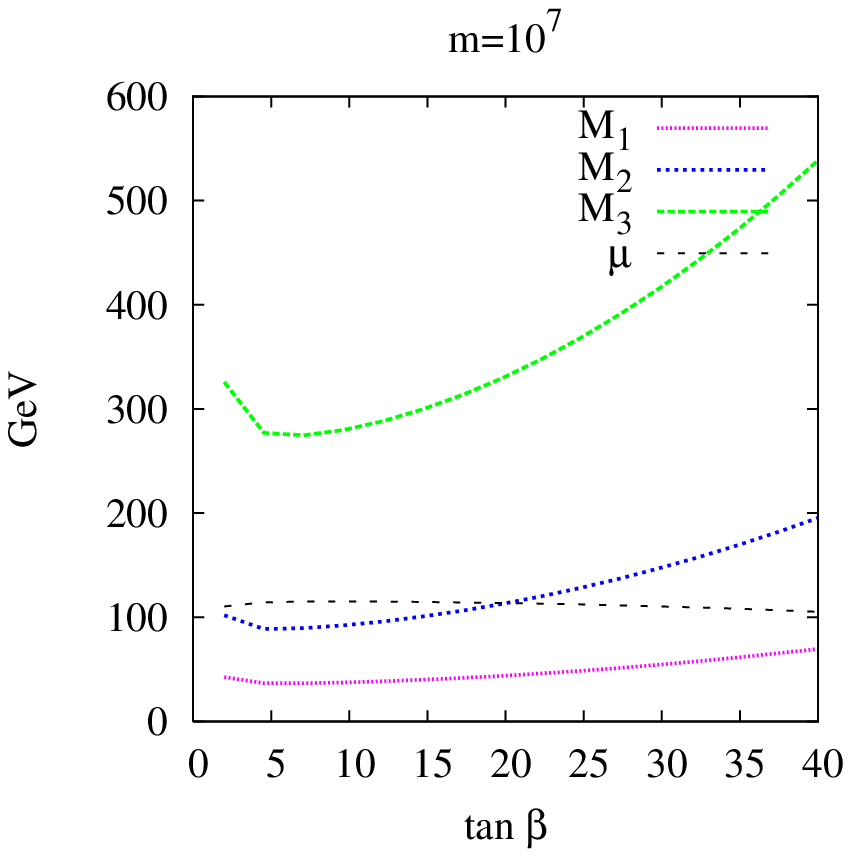}
{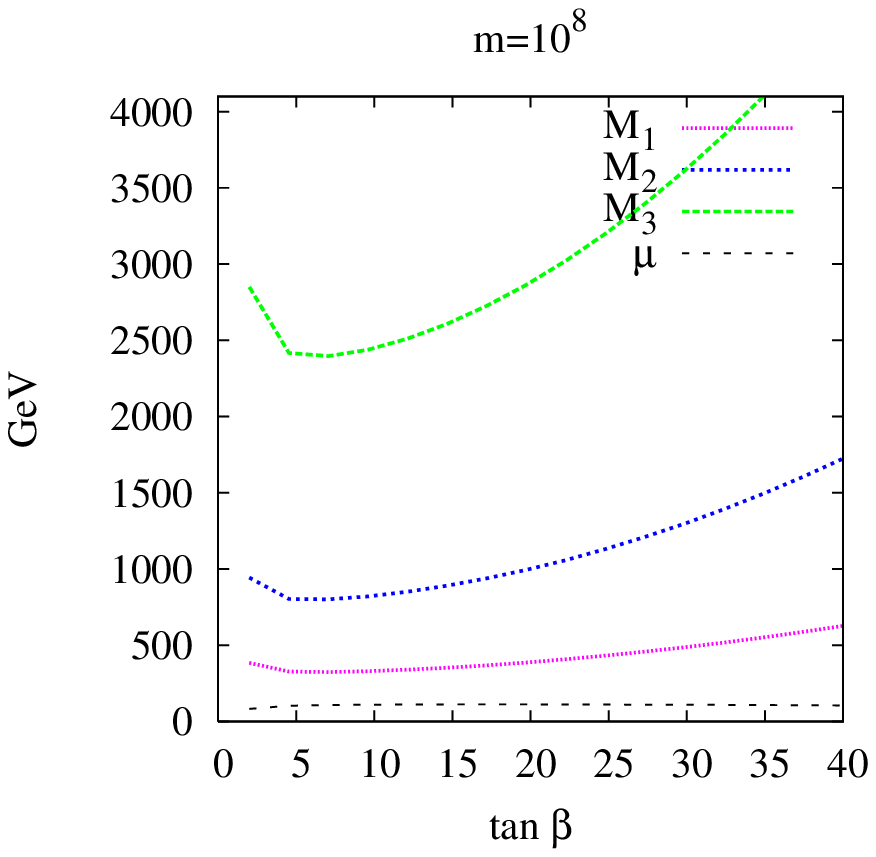}}
\caption{Gaugino masses and the $\mu$-term for $\tilde{m} = 10^6,
10^7, 10^8$GeV. The boundary condition at the string scale is 
$M_1 = M_2 = M_3 = 1$GeV and $\mu = 100$GeV.\label{splitGUTunifig2}}
\end{figure}

Next we study the RGE evolution of gaugino masses and the $\mu$-term. We assume
that the gaugino masses at the high scale are close to $0$GeV and
that the magnitude of the $\mu$-term is tuned to $100$GeV.
The A-terms are taken to be negative and between $10^4$ and $10^7$ GeV in
magnitude. With these initial conditions at the string scale $m_s$ we
evolve down to the scalar mass scale $\tilde{m}.$
The results for gaugino masses and the $\mu$-term at the scale of 
$M_Z$ are shown in Figure \ref{splitGUTunifig2}.
The salient features are a heavy gluino, a slight change
in the $\mu$-term, a small increase in $M_1$, and a moderate one in $M_2.$


We have checked that the results are largely insensitive to the
choice of gaugino masses at $m_s$ in the range $0-10$GeV.
We also investigated the effects of modifying the RGE equations above 
the scale $\tilde{m}$ so that the gauge couplings unify at the string 
scale of around $10^{13}$GeV. 
This is achieved by having 2 extra lepton doublets and 3 extra lepton
singlets. Again, the results for gaugino masses and the $\mu$ term
at scale $M_Z$ are largely unaffected.



The next issue to address is that of dark matter in this scenario.
As discussed in \cite{hepph0406088}, there are three
different possibilities. 

The Bino by itself cannot be the dark matter particle, since it is
a gauge singlet and only interacts with the Higgs and higgsinos through
the terms
\begin{equation}
{H^\dag\over\sqrt{2}} \tilde{g}'_u \tilde{B} \tilde{H}_u +
{H^T \epsilon\over\sqrt{2}} \tilde{g}'_d \tilde{B} \tilde{H}_d + {\rm h.c.}
\end{equation}
Therefore not a sufficient number of channels are available for its
annihilation and the resulting dark matter density would be too large 
\cite{hepph0406088}.

The LSP could be a mixture of Bino
and higgsino if $M_1$ and $\mu$ are comparable in size, with relic density
\begin{equation}
\label{darkmattereqn}
\Omega_\chi h^2 \approx 0.1 {{{\mu^2} (M_1^2 + \mu^2)^2}\over{m_\chi^4
  {\rm TeV}^2}}.
\end{equation}

\begin{figure}
\includegraphics{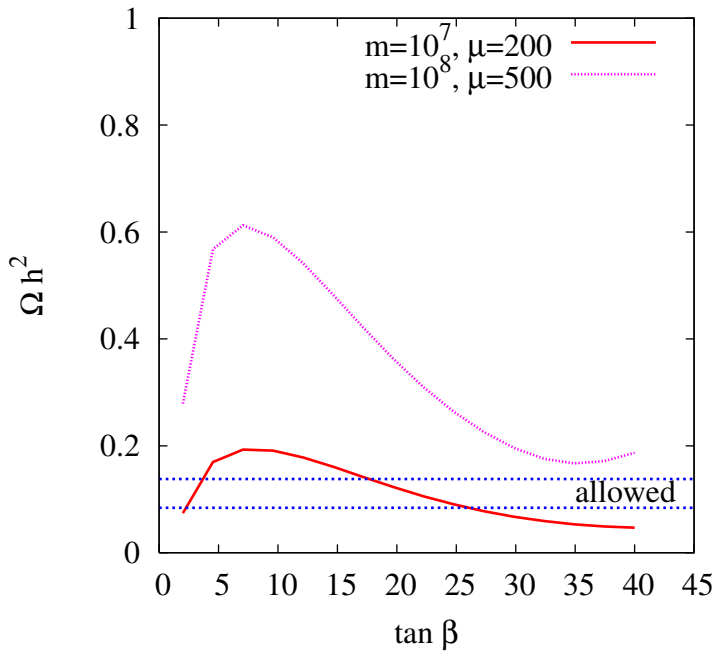}
\caption{The dark matter relic density $\Omega h^2$ in the heavy scalar 
scenario, for $\tilde{m}=10^7$GeV, $\mu=200$GeV and $\tilde{m}=10^8$GeV,
$\mu=500$GeV.}
\label{darkmatter}
\end{figure}

We consider the $\tilde{m}=10^7$GeV scenario, with $\mu=200$GeV at the
string scale. The relic density is computed using formula (\ref{darkmattereqn})
and its dependence on $\tan\beta$ is shown in
Figure \ref{darkmatter}. Also included is the graph of the 
relic density for
$\tilde{m}=10^8$GeV and $\mu=500$GeV at the string scale. It is clear
that with appropriately chosen values of $\mu$ and $\tan\beta$ 
it is possible to obtain an acceptable dark matter relic density.
The particle spectrum for $\tilde{m}=10^7$GeV, $\mu=200$GeV and
$\tan\beta=22$,
satisfying the dark matter hypothesis, is shown in Table \ref{splitsusyspectrum}.

\begin{table}
\begin{center}
\begin{tabular}{|c|c|}
\hline
Particle & Mass \\
\hline
$\chi_1^0$ & 123\\
\hline
$\chi_2^0$ & 204\\
\hline
$\chi_3^0$ & 240\\
\hline
$\chi_4^0$ & 407\\
\hline
$\chi_1^+$ & 212\\
\hline
$\chi_2^+$ & 383\\
\hline
$\tilde{g}$ & 1033\\
\hline
$\Omega h^2$ & 0.11\\
\hline
\end{tabular}
\caption{Supersymmetric particle spectrum for $\tilde{m}=10^7$GeV, $\mu=200$GeV and $\tan\beta=22.$
\label{splitsusyspectrum}}
\end{center}
\end{table}

The second case is that of a heavy higgsino being the LSP. In this case
the relic abundance is
\begin{equation}
\Omega_{\tilde{H}} h^2 = 0.09 \left( \mu\over{\rm TeV} \right)^2.
\end{equation}

Scenarios with $\tilde{m}$ slightly above $10^8$GeV can realise this 
possibility, since then the Bino has mass above $700$GeV. With a $\mu$-term
chosen to be $1-1.2$TeV at the string scale $m_s$, it is possible to 
obtain the correct relic abundance. For example, with $A=-3\times 10^6$GeV and
$\tan\beta=40$ we have (at scale $M_Z$) $M_1\sim 2200$GeV, $M_2\sim
5600$GeV, $M_3\sim 12000$GeV whereas $\mu=1006$GeV.

The last possibility is that of a Wino LSP and relic abundance
\begin{equation}
\Omega_{\tilde{W}} h^2 = 0.02 \left( {M_2\over{\rm TeV}} \right)^2.
\end{equation}
However, this cannot be realised in our scenarios, since the Wino is 
always heavier than the Bino.

\subsection{Stringy mSUGRA}
We note that the string scale is independent of the flux superpotential
parameter $W_0$, whereas all the soft terms are directly proportional
to it. Therefore we can take $m_s \sim 10^{16}$GeV, put all the matter
on D3 branes and lower the value of $W_0$ to obtain realistic 
values for the scalar masses and other soft parameters. Note that
the gravitino mass will also be lowered in this process, and its
size will be roughly just one order of magnitude above the scalar
masses.
We must be careful for a number of reasons, since lowering $W_0$ means
lowering the magnitude of supersymmetry breaking in the nonSUSY
AdS minimum of \cite{hepth0505076}.
Therefore the D-term contribution to SUSY breaking might become significant
compared to the F-term contributions. However, this is not the case
since both D and F-terms are proportional to the value of $W_0.$

Therefore, we have two parameters to control - $\cal{V}$ and $W_0.$
Increasing $\cal{V}$ decreases the string scale, while decreasing
$W_0$ decreases the soft terms only (since the amount of supersymmetry breaking
is reduced). However, the only reasonable scenarios with scalar masses
of ${\cal{O}} (1 {\rm TeV})$ will arise from the scenario with string scale
$10^{16}$GeV and appropriately reduced $W_0.$ The string scale cannot
be reduced any further, since that would entail increasing ${\cal{V}}$ 
significantly and the ratio of gaugino to scalar masses would become very
small, leading back to the split SUSY scenario.

\section{Other possibilities}

One could also consider D3 or D7 branes in a warped region, so that
the warp factor reduces the scalar masses to a small value. 
This however also reduces the value of the string scale to a smaller value.
The resulting small gap between the scalar mass scale and the string scale
makes it difficult to achieve unification at the string scale - one needs
a large amount of extra matter. For example, if the string scale
is lowered to $10^6$ GeV, one needs at least 14 new multiplets,
and with the string scale at $10^4$GeV, one needs 51. The
 construction of  string models in which this type of scenarios can be
realised is open.

\section{Conclusions}

We have presented three distinctive scenarios of low-energy
supersymmetry breaking which were derived from string
compactifications.
It is very appealing to be able go such a long way - from a string theory
compactification with fluxes to computing the spectrum at
low energy of specific models. It is also illustrative to see how
difficult it is to obtain fully realistic scenarios given the many
cosmological and phenomenological constraints.

The three scenarios we studied have several interesting properties and
represent a progress in the right direction to close the gap between
string theory and low-energy physics. In particular the detailed spectrum
of supersymmetric particles was obtained at low-energies for
representative examples. 

Each of the scenarios has  potential problems which illustrate the
difficulties that can be expected in general. The generalised fluxed MSSM
scenario, despite its interesting features, has to face the
cosmological moduli problem which appears at every value of the string
scale, as mentioned in the text. We would have to rely on a solution
of this problem in
order to consider this scenario to be fully realistic. Notice that
this potential problem could not have been envisioned in the original
discussion of soft supersymmetry breaking induced by fluxes since there
was no mechanism to fix the K\"ahler structure moduli. 
Also, for scales lower than $10^{12}$GeV the stop is the LSP instead
of a neutralino. Barring these
problems we found that the parameter space is greatly reduced by the
standard phenomenological constraints. The case $\mu>0$ is
not compatible with the boundary condition for
the $B$ term, while for $\mu<0, m_s<10^{16}$GeV only low $\tan\beta$
solutions exist and the Higgs mass is below the LEP2 bound.
This problem may be alleviated by relaxing the $B$ term boundary
condition, e.g. by considering the Higgs coming from D3-D7 strings. Our lack of
control of the K\"ahler potential for these fields does not allow us to make a
concrete statement about this possibility.

The intermediate scale split supersymmetry scenario needs the usual
fine tuning in order to keep the Higgs light. Again the
D3-D7 particles could alleviate this problem if their masses were
under control. Furthermore, despite sharing with the
split supersymmetry scenario the property that the scalars are hierarchically
heavier than the gauginos, it has to be pointed out that in our
scenario the string scale is intermediate and therefore we do not
expect the standard MSSM gauge coupling unification to work.

The stringy mSUGRA scenario requires a tuning of the flux superpotential
in order to achieve a GUT string scale. This tuning however may be argued 
to be less severe
than the tuning of the split supersymmetry that needs to be made at
the spectrum level, although both rely on the existence of many flux
vacua.
Notice that in the stringy mSUGRA scenario, without knowing any relationship
among the relative coefficients,  we can only estimate the order of
magnitude of the soft breaking terms. 

It is fair to say also that each of the phenomenological constraints
we have used may be relaxed and the parameter space of realistic
models can be substantially enhanced. 
For example, non-universal flavour structure in the soft SUSY breaking
terms can cancel the supersymmetric contribution to BR$(b\rightarrow s
\gamma)$. Also, evidence for a non-Standard Model component of $(g-2)_\mu$ is
controversial and may turn out to not be relevant. Similarly, dark matter
could all reside in the hidden sector and the LSP could be unstable.

It is interesting to compare our results with the phenomenology of
KKLT-type scenarios. The essential difference is that the large volume
minima we found are non-supersymmetric even before the lifting term
is added. In KKLT scenarios, supersymmetry is restored after fixing
the K\" ahler moduli, so it is precisely the anti D3 brane term which is
responsible for supersymmetry breaking. Of course, the uplift
gives non-zero values to F-terms as well, but these will be
parametrised in terms of the uplift potential.
In fact it turns out \cite{hepth0503216,hepph0507110} that they are
significantly suppressed with respect to the no-scale model
values. This implies that, for example, the moduli mediated
contribution to the D7 scalar masses is suppressed with respect
to $m_{3/2}$ and will generically compete with the AMSB contribution.
The same is true for gaugino masses, where it turns out that it is
possible to obtain a `mirage unification' at the intermediate
or even TeV scale by combining the two contributions 
\cite{hepph0504037,hepph0504036, hepph0511320}. There were no
realistic scenarios for D3 soft terms in that scenario.

Finally, it is worth pointing out that what we have analysed here are
general scenarios assuming that the Standard Model is on D3 or D7
branes. In order to be more concrete it would be desirable to carry out
this kind of analysis on explicit D-brane models, taking into account
potential matter and gauge fields beyond the MSSM, etc. There are clearly
many things to be done in this direction. Explicit model building
may become more focused after data from the Large Hadron 
Collider (LHC) arrives.

\section*{Acknowledgments}
We would like to thank Cliff Burgess, Joseph Conlon
and Luis Ib\'a\~nez for useful conversations. We also thank 
Gordon Kane and Piyush Kumar
for comments on an earlier version of the paper. This research is
partially funded by PPARC. FQ is
also funded by a Royal Society Wolfson award. KS is grateful to
Trinity College, Cambridge for financial support.

\begin{appendix}
\section{Appendix: Gaugino Masses in No-Scale Models}

Let us consider the general
formula for the gaugino mass in models with both moduli and anomaly mediated
SUSY breaking, as derived in \cite{hepth9911029}:
\begin{equation}
\label{amsbgaugino}
m_{G} = - {g^2\over{16\pi^2}} \left( (3T_G-T_R) m_{3/2} + (T_G-T_R) K_i F^i
+ {2T_R\over d_R} (\log\det K|_R^{''})_{,i} F^i \right)
\end{equation}
Here $T_G$ is the Dynkin index for the adjoint representation, $3T_G-T_R$
is the one-loop gauge beta function coefficient and $T_R$ is the Dynkin index associated with 
the representation $R$ of dimension $d_R$, equal to $1/2$ for the
$SU(N)$ fundamental; a sum over all the matter representations $R$ is
understood in each term with the $R$ subindex.
 $K$ is the K\" ahler potential, $K''|_R$ its
second derivatives with respect to matter fields projected onto the
corresponding representation of $G$, and the F-terms
are defined as\footnote{The minus sign is inserted to make the
definition agree with the global supersymmetry one in the limit 
$M_P\to\infty.$}
\begin{equation}
F^{i} = - e^{K/2} K^{i\bar{\jmath}}\,\overline{D_j W},
\end{equation}
with $W$ the superpotential.
The first term in equation (\ref{amsbgaugino}) is the usual super-Weyl anomaly
induced term present in any supergravity theory. The second and third terms
arise from K\" ahler and sigma-model anomalies.

For simplicity let us consider the case of a single K\" ahler modulus
model. Let us assume that the K\" ahler potential is of the no-scale form,
\begin{equation}
K = -3 \log \left( T+T^* - {1\over3} \phi \phi^* \right)
\end{equation}
where $\phi$ is the matter field. Taking derivatives with respect
to matter fields, one gets the result for a particular representation $R$
\begin{equation}
{1\over{d_R}} \log \det K|_R^{''} = {1\over3} K,
\end{equation}
assuming that the vevs of visible matter vanish.
Hence from formula (\ref{amsbgaugino}) we get 
\begin{eqnarray}
\label{gaugmass}
m_{G} &=& 
 - {g^2\over{16\pi^2}} \left( (3T_G-T_R) m_{3/2} + (T_G-T_R) K_i F^i
 + {2\over3} T_R K_i F^i \right)\\  &=&
  - {g^2\over{48\pi^2}} (3T_G - T_R) (3m_{3/2} + K_i F^i).
\end{eqnarray}
In no-scale models SUSY breaking corresponds to nonzero values of the 
auxiliary field corresponding to $T.$ The gravitino mass is equal to
\begin{equation}
m_{3/2} = e^{K/2} W.
\end{equation}
Also
\begin{equation}
F^T = - e^{K/2} K^{T\bar{T}} D_T W = 
- e^{K/2} {(T+T^*)^2\over3} (\partial_T K)W
    = e^{K/2} (T+T^*) W.
\end{equation}
assuming that $T$ is absent from the superpotential.
Therefore we have
\begin{equation}
K_T F^T + 3 m_{3/2} = - 3 e^{K/2} W + 3 e^{K/2} W = 0.
\end{equation}
It can easily be checked that this result generalises to the $n$-K\" ahler
modulus case, due to the no-scale property $K^{i{\bar{\jmath}}} \partial_i K
\partial_{\bar{\jmath}} K = 3$ with $i,j$ ranging over all K\" ahler moduli.
It follows from equation (\ref{gaugmass}) that the AMSB contribution
to gaugino masses is vanishing.

\end{appendix}

\end{document}